\documentclass[10pt,twocolumn,letterpaper]{article}

\pdfoutput=1

\usepackage[pagenumbers]{cvpr} 

\usepackage{graphicx}
\usepackage{amsmath}
\usepackage{amssymb}
\usepackage{booktabs}

\usepackage[pagebackref,breaklinks,colorlinks]{hyperref}

\usepackage[capitalize]{cleveref}
\crefname{section}{Sec.}{Secs.}
\Crefname{section}{Section}{Sections}
\Crefname{table}{Table}{Tables}
\crefname{table}{Tab.}{Tabs.}

\def\cvprPaperID{10580} 
\def\confName{CVPR}
\def\confYear{2022}

\usepackage{amsmath}
\usepackage{amssymb}

\newcommand{\orig}{x}
\newcommand{\pred}{\tilde{x}}
\newcommand{\reco}{\hat{x}}

\newcommand{\latentC}{\tilde{y}}
\newcommand{\res}{r}
\newcommand{\genRes}{g}
\newcommand{\xD}{\hat{x}_\mathrm{d}}
\newcommand{\xG}{\hat{x}_\mathrm{g}}
\newcommand{\thresh}{\vartheta_\mathrm{t}}

\newcommand{\encoderNet}[3]{f_a[#1,#2,#3]}
\newcommand{\decoderNet}[3]{f_s[#1,#2,#3]}
\newcommand{\encoderPNet}[3]{h_a[#1,#2,#3]}
\newcommand{\decoderPNet}[3]{h_s[#1,#2,#3]}
\newcommand{\GDNet}[3]{\operatorname{GD}[#1,#2,#3]}
\newcommand{\GSNet}[3]{\operatorname{GS}[#1,#2,#3]}
\newcommand{\hypCoderNet}[3]{\operatorname{P}[#1,#2,#3]}
\newcommand{\condCoderNet}[2]{\operatorname{CAE}[#1,#2]}
\newcommand{\resCoderNet}[1]{\operatorname{RAE}[#1]}
\begin{document}

\title{Generalized Difference Coder: A Novel\\ Conditional Autoencoder Structure for Video Compression}

\author{Fabian Brand, J\"urgen Seiler, Andr\'e Kaup\\
Multimedia Communications and Signal Processing, \\Friedrich-Alexander-Universit\"at Erlangen-N\"urnberg\\
Cauerstr. 7, 91058 Erlangen, Germany\\
{\tt\small \{fabian.brand,juergen.seiler,andre.kaup\}@fau.de}
}
\maketitle

\begin{abstract}
   Motion compensated inter prediction is a common component of all video coders. The concept was established in traditional hybrid coding and successfully transferred to learning-based video compression. To compress the residual signal after prediction, usually the difference of the two signals is compressed using a standard autoencoder. However, information theory tells us that a general conditional coder is more efficient. In this paper, we provide a solid foundation based on information theory and Shannon entropy to show the potentials but also the limits of conditional coding. Building on those results, we then propose the generalized difference coder, a special case of a conditional coder designed to avoid limiting bottlenecks. With this coder, we are able to achieve average rate savings of 27.8\% compared to a standard autoencoder, by only adding a moderate complexity overhead of less than 7\%.
\end{abstract}

\section{Introduction}
The invention of autoencoders for image compression~\cite{BalleLS2017_Endendoptimized} has shaped the research in this area decisively for the last few years. Similar to existing image and video compression standards like JPEG~\cite{Wallace1992_JPEGstillpicture}, JPEG2000~\cite{ITUTI2004_JPEG2000Image}, HEVC~\cite{SullivanOH2012_OverviewHighEfficiency}, VVC~\cite{BrossCL2020_VersatileVideoCoding}, AV1~\cite{HanLM2021_TechnicalOverviewAV1}, or many more, this method relies on a transformation of the image into a sparse domain. Different to the traditional methods, transforms in learning-based compression are non-linear and data-driven. Typically, these transforms are implemented as convolutional neural networks which are trained as a constrained autoencoder. 

Learning-based approaches have also been proposed for video compression. The general structure of the approaches has largely been taken over from known hybrid video coders such as VVC or AV1. At first, motion is estimated, then the motion field is transmitted (typically in a lossy way). Afterwards motion compensation is performed. This yields a prediction frame which is used to reduce the temporal redundancy between frames. The task at hand is now to transmit the remaining information to obtain the reconstructed frame. In most published learning-based approaches~\cite{LuOX2018_DVCEndend,LuCZ2020_ContentAdaptiveError,HuCX2020_ImprovingDeepVideo}, the strategy is taken over from traditional video compression, i.e., the difference between the original frame and the prediction frame, the so-called residual signal, is computed and transmitted. At the decoder, the reconstructed residual is added to the prediction signal to obtain the reconstructed frame. This strategy has been proven efficient in both traditional video coding and also in deep-learning-based video coding. However, in the latter case, the neural-network-based structure allows for a different possibility: Conditional coding. In conditional coding, we do not compress the residual, but rather the frame itself under the condition of knowing a prediction. We can therefore exploit a relationship from information theory: The entropy of the difference is greater or equal the conditional entropy of the signal $\orig$ given its prediction $\pred$:
\begin{equation}
H(\orig - \pred) \ge H(\orig|\pred)
\end{equation}
This inequality suggests that in theory conditional coding is always at least as good as conventional residual coding. Using neural networks, conditional coding can be implemented as a conditional autoencoder. The first use of a conditional autoencoder in the context of image and video compression was proposed in~\cite{BrandSK2019_IntraFramePrediction} for intra prediction.

In this work, we thoroughly examine the possibility of conditional coding for video compression. At first, we show the potential but also the limits of such an approach on the basis of information theoretical considerations. Based on these results, we then construct a novel approach, called the generalized difference coder (GDC), a conditional autoencoder designed to minimize the shortcomings of conditional coders with very small computational overhead. We extend this approach further to gain more robustness in cases with very small residual by additionally coding the residual in a conditional approach. Finally, we demonstrate the effectiveness of our proposed methods in various experiments.

\section{Related Work}
All modern video coders make use of inter prediction to reduce the temporal redundancy between frames. Block-based coders such as HEVC~\cite{SullivanOH2012_OverviewHighEfficiency}, VVC~\cite{BrossCL2020_VersatileVideoCoding}, VP9~\cite{MukherjeeBG2013_latestopensource} or AV1~\cite{HanLM2021_TechnicalOverviewAV1} estimate motion on a block-level, yielding a prediction signal per block. If no suitable block is available in the reference frame, the coders have the possibility to locally switch back to intra prediction using previously decoded content. After prediction, a residual is computed by subtracting the prediction signal from the original. The resulting residual is then transformed using a handcrafted frequency transform such as the discrete cosine transform, a discrete sine transform or a combination thereof. Furthermore, AV1 and VVC allow dynamic switching between different transforms.

In 2019, the deep video compression framework (DVC)~\cite{LuOX2018_DVCEndend} was published. This work introduced the first end-to-end trained video coder consisting of motion estimation, motion transmission, and residual transmission. Additionally, DVC contained a network to enhance the prediction frame. For motion estimation, a SpyNet~\cite{RanjanB2017_OpticalFlowEstimation} architecture is used. Both motion and residual compression use a standard autoencoder, similar to~\cite{BalleLS2017_Endendoptimized}. 

With DVC it was possible to compress images with one given reference picture. However, in practical applications, it is often beneficial to use multiple reference pictures. This approach helps avoiding occlusions in the predicted frame. One example where multiple reference frames were used is \cite{LinLL2020_MLVCMultiple}, where the authors proposed a framework which transmits multiple motion vector fields using motion vector prediction and generate a joint prediction frame. This prediction frame is again used to compute a residual which is then transmitted.

\begin{figure}
	\centering
	\begin{tabular}{cc}
		\includegraphics[height=0.6\columnwidth]{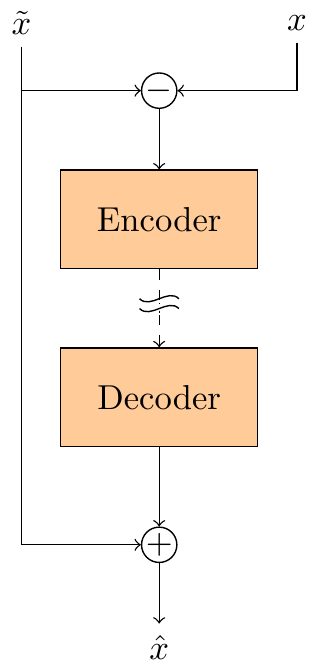} &
		\includegraphics[height=0.6\columnwidth]{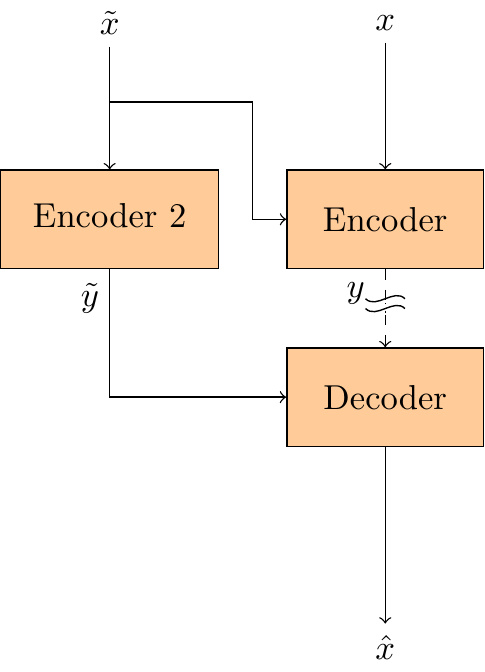}\\
		Difference Coder & Conditional Coder from\\& \cite{LadunePH2020_ModeNetModeSelection} and \cite{LadunePH2020_OpticalFlowMode}\\
	\end{tabular}
	\caption{Left: Conventional residual coder. Right: Schematic network structure of conditional autoencoder as proposed as \textit{CodecNet} in \cite{LadunePH2020_ModeNetModeSelection} and \cite{LadunePH2020_OpticalFlowMode}. Dashed lines denote transmission over a channel.}\label{Fig:SchematicsLadune}\vspace{-0.5cm}
\end{figure}

The previously mentioned frameworks all follow the basic structure taken from traditional inter coding approaches. In \cite{LadunePH2020_ModeNetModeSelection}, Ladune \etal proposed CodecNet, a conditional coding approach, together with ModeNet, which enables a skip mode. Skip modes are commonly found in traditional codecs and describe copying of the prediction signal directly into the reconstructed frame without residual transmission. As shown in Fig. \ref{Fig:SchematicsLadune}, the conditional coding approach used in CodecNet consists of two encoders. One conditional branch encoder which generates a latent representation of the prediction signal and a main encoder, which gets both the prediction signal and the original signal. The encoder therefore compresses the original signal under the condition of knowing the prediction signal. The resulting latent representation is coded and transmitted over the channel. The decoder uses the compressed latent representation of the original and the latent representation of the prediction signal to reconstruct the original frame. Note that the latent representation of the prediction signal is not transmitted since it can be constructed at the decoder. The decoder therefore reconstructs the frame under the condition of knowing \emph{a latent representation} of the prediction signal.

The same basic structure was used in \cite{LadunePH2020_OpticalFlowMode}, where the authors proposed a joint estimation and transmission of motion vector field and skip mode. In a further publication~\cite{LadunePH2021_ConditionalCodingVariable}, Ladune proposed a framework for multi-reference-frame coding using a similar paradigm of conditional coding.

\section{Foundations in Information Theory}
\label{Sec:Theory}
In order to get a better understanding of residual coding in video compression, we want to consider the different scenarios in the light of information theory. The Shannon entropy serves as a lower bound of the bitrate needed to compress a signal. In this section, we assume ideal lossless coders which reach entropy. Let $H(x)$ denote the entropy of the distribution from which $x$ is drawn. I.e., if $x$ symbolizes a natural image, $H(x)$ is the entropy of natural images. In residual coding, we compress the residual $r=\orig-\pred$ and in conditional coding, we compress the original frame $\orig$ under the condition of knowing $\pred$. We therefore have to compare the entropy of residual frames $H(r) = H(\orig - \pred)$ and the conditional entropy $H(\orig|\pred)$.

We start by using Bayes law, to derive
\begin{equation}
	H(\orig,\pred|\res) + H(\res) = H(\orig,\pred,\res) = H(\orig,\pred) + \underbrace{H(\res|\orig,\pred)}_{=0}.
\end{equation}
We can easily see that the conditional entropy $H(\res|\orig,\pred)$ must be zero, since the residual is completely determined from knowing $\orig$ and $\pred$.
We can therefore continue by summarizing and rearranging:
\begin{equation}
\begin{split}
H(\res) &= H(\orig,\pred) - H(\orig,\pred|\res) \\
&=H(\orig|\pred) + H(\pred) - \underbrace{H(\orig|\pred,\res)}_{=0} - H(\pred|\res) \\
&=H(\orig|\pred) + H(\pred) - H(\pred|\res) = H(\orig|\pred) + I(\pred,\res).
\end{split}
\end{equation}
In the second line, we can see that $H(\orig|\pred,\res)$ must be zero because the original frame $\orig$ can be reconstructed from the prediction signal $\pred$ and the residual $\res$. $H(\pred) - H(\pred|\res)$ is the mutual information $I(\pred,\res)$ and so we can write:
\begin{equation}
	H(\orig-\pred) = H(\orig|\pred) + I(\pred;\res)
	\label{Eq:Main}
\end{equation}
Since the mutual information is non-negative, this implies
\begin{equation}
H(\orig-\pred) \geq H(\orig|\pred),
\end{equation}
with equality if and only if $\res$ and $\pred$ have no mutual information.

From this inequality we obtain general insights about the efficiency of conditional coding compared to residual coding. We not only see that conditional coding is (in theory) at least as good as residual coding but we can also quantify how large the difference is. The larger the mutual information between the residual and the prediction frame, the larger is the gain of conditional coding. 

When interpreting the results, we need to take into account possible bottlenecks between the prediction signal and the output. One example where such a bottleneck appears is the conditional autoencoder proposed in \cite{LadunePH2020_ModeNetModeSelection} and \cite{LadunePH2020_OpticalFlowMode}. The schematic structure of this coder is given in Fig~\ref{Fig:SchematicsLadune}. Here, an additional encoder is used to obtain a latent representation of the prediction signal, which is then used together with the transmitted latent representation to reconstruct the frame. The decoder only sees a latent representation of the prediction signal. Let $\latentC = f\left(\tilde{x}\right)$ be that latent representation. It is clear that
\begin{equation}
	H(\pred) \ge H(\latentC)
	\label{Eq:Entropy}
\end{equation}
and
\begin{equation}
	H(\orig|\pred) \le H(\orig|\latentC)
	\label{Eq:CondEntropy}
\end{equation}
hold. Furthermore, it can be shown that 
\begin{equation}
H(\orig|\pred) = H(\orig|\latentC) - I(\orig;\pred|\latentC)
\label{Eq:CondEntropy2}
\end{equation}
Plugging this result into (\ref{Eq:Main}), we obtain
\begin{equation}
H(\orig-\pred) = H(\orig|\latentC) - I(\orig;\pred|\latentC) + I(\pred;\res).
\label{Eq:Main2}
\end{equation}
This shows, that $H(\orig-\pred)$ is not necessarily smaller than $H(\orig|\latentC)$, depending on how much information is lost during the transform $f$. 

We want to illustrate the relations on an example. Let us consider the extreme case $\orig = \pred$. In this case, $I(\pred;\res) = 0$ and we see from (\ref{Eq:Main}) that $H(\orig-\pred) = H(\orig|\pred)$. We can conclude that the ideal conditional autoencoder can not be better than the residual coder in this case. Moreover, when the conditional path contains a bottleneck as proposed in \cite{LadunePH2020_ModeNetModeSelection}, we can follow from (\ref{Eq:Main2}) and from $I(\orig;\pred|\latentC) > 0$\footnote{$I(\orig;\pred|\latentC)=0$ only holds when no bottleneck is present.} that in this case a real conditional autoencoder will perform worse. For $\orig = \pred$, we can follow that $I(\orig;\pred|\latentC)\big|_{\orig = \pred} = H(\pred|\latentC)$. So the performance of the conditional approach becomes worse when the transform $f$ removes information. In other words: When the prediction signal is too good, the overall bottleneck of the coder shifts to the conditional branch, which is therefore an upper limit for the reconstruction quality. In a residual coder, this problem does not exist, since the prediction signal is passed to the decoder without any processing. 

We see that in the absence of a bottleneck, e.g., when $\latentC = \pred$, we obtain
\begin{equation}
	I(\orig;\pred|\latentC)\bigg|_{\latentC = \pred} = I(\orig;\pred|\pred) = 0
\end{equation}

Note that these deliberations consider any kind of bottleneck. One possible bottleneck is the classical bottleneck occurring in many autoencoders, where the dimensionality of the latent space is lower then the dimensionality of the input space. This is for example the case in \cite{LadunePH2020_OpticalFlowMode}. Another kind of bottleneck arises from the general fact that
\begin{equation}
	H(\pred) \ge H(f(\pred))
\end{equation}
This shows that no function can increase the entropy of a signal, a result which is also known as the data processing theorem~\cite{ViterbiO1979_PrinciplesDigitalCommunication}. In reality functions which fulfill this equation with equality are rare and usually hand-crafted, such as for example the discrete cosine transform, which is often found in image and video coder and which is a reversible transform. In particular, convolutions are not generally reversible. This indicates that each layer of a neural network, which processes the prediction frame, may reduce its entropy and therefore decrease the upper bound for the reconstruction quality of the network. From these observations we conclude that the performance of conditional autoencoders can be improved when avoiding bottlenecks in the conditional branch.

\section{Generalized Difference Coder}
\begin{figure}
	\centering
	\begin{tabular}{cc}
		\includegraphics[width=0.3\columnwidth]{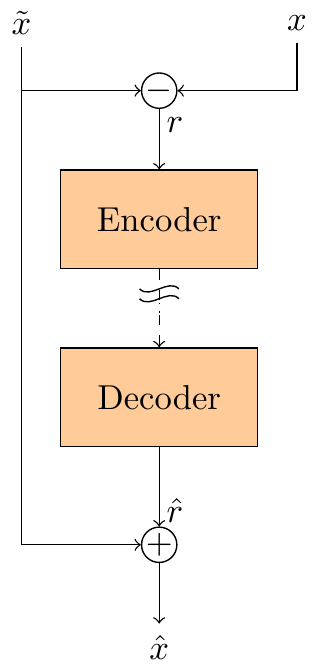} &
		\includegraphics[width=0.3\columnwidth]{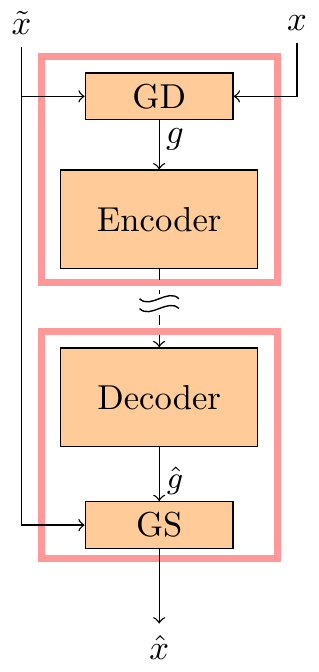}\\
		Difference Coder & Generalized Difference Coder\\
	\end{tabular}
\caption{High-level schematics of difference coder and generalized difference coder.}\label{Fig:Schematics}\vspace{-0.5cm}
\end{figure}

With the deliberations from the previous section in mind, we propose the generalized difference coder (GDC). At the core of this approach are two small sub-networks, which we call generalized difference (GD) and generalized sum (GS). These networks take the place of the difference and sum of a conventional residual coder, respectively. 
In Fig. \ref{Fig:Schematics}, we give a schematic comparison between the conventional residual coder (in the following called ``difference coder'' for better distinction) and the proposed generalized difference coder GDC. When we look at GD and the encoder and at GS and the decoder, we see that this structure matches the general structure of a conditional autoencoder, as indicated by the red frames in the figure. 
Comparing our proposal to CodecNet from \cite{LadunePH2020_OpticalFlowMode}, we were able to remove the encoder in the conditional branch. This has two advantages: First, removing the additional encoder from the prediction path also removes the bottleneck following this encoder. As derived in the previous section, this bottleneck limits the possible performance of the overall network.

Second, the additional encoder network would have to be run during decoding time, thus increasing the decoding time. The additional encoder has about the same size as the decoder. Thereby, the structure from \cite{LadunePH2020_OpticalFlowMode} doubles the decoder complexity compared to a difference coder. The networks in our approach have a much smaller complexity.

We now want to formally define the GD and GS operators. We recall the properties of the difference: A difference has two inputs and one output, all of which are of the same size, both in the spatial dimension and in the number of color components. Furthermore, the difference is a linear and a local operator. The GD also has two inputs, but the output may have a different number of channels (usually larger) than the two inputs, which are of the same size. The spatial dimensions remain untouched. Also, GD is non-linear and the result of a pixel may be influenced by the local environment via convolutional layers. Analogously to the residual, which is the output of the difference operator, we denote the output of the GD operator the \emph{generalized residual} $\genRes$. The GS operator has similar properties. The main difference lies in the  dimensionality of input and output. Other than for GD, here, the output has the same dimensionality as one of the inputs, while the other (the one originating from the GD) may have a different number of channels but the same spatial size.

Our proposed structure also has the advantage of being very flexible and generic. The autoencoder at the core of the network is completely exchangeable, so the GDC structure is independent of the autoencoder architecture which was chosen. Also the choice of the networks representing GD and GS can be selected freely under the above constraints. 

\section{Switchable Residual Compression}
In preliminary experiments, we found that even though the GDC does not have a bottleneck between prediction and reconstruction, it still does not outperform the difference coder for all types of images. Again, this particularly concerns images with very small $I(\pred;\res)$. Even though the problems are not severe, these images reduce the overall performance of the GDC. In such cases in compression it is always worth to examine the possibility of hybridization. However, it turns out that hybridization of GDC and difference coder are not easily possible. The reason is that deep-learning-based coders typically process the frame as a whole or in very large blocks. This is due to the successive downsampling of the frame to the latent space and often further in a hyperprior network. The smallest possible tile which can be independently decoded is therefore very large (in typical structures at least $64\!\times\!64$). If we want to switch on a granularity smaller than 64 pixels, we would have to fully transmit both possibilities. This strategy is not efficient, since twice the rate is needed.

To solve this problem, we propose to \emph{jointly} encode the generalized difference $\genRes$ and the linear residual $\res$ into one latent space. That way, we can decode both frames from one latent space. This enables switching on arbitrary granularities, since the decoder can decode both representations at the same time. At the same time, the autoencoder can exploit all redundancy which exists between the generalized residual and the linear residual. At the decoder, we decode thus two different reconstructed frames:
\begin{equation}
	\xD = \pred + \hat{\res}
\end{equation}
and 
\begin{equation}
	\xG = \operatorname{GS}\left(\pred, \hat{\genRes}\right).
\end{equation}
Which signal shall be used where is to be determined by the decoder and transmitted as side-information.

This strategy also has other advantages. For one, the core autoencoder network directly knows the (linear) residual. It is commonly known that coding the difference between original and prediction is an efficient residual coding technique. By providing the autoencoder with this signal, we can assure that a valid and sensible representation is present from the beginning. That way, the autoencoder can converge more stably since part of the input does not depend on another network. Furthermore, GD is not burdened with finding a signal with at least as much information as the residual and can instead focus on extending this information for more efficient encoding. Similarly, we can argue that the possibility to reconstruct the frame without the involvement of a neural network supports the initial convergence. 

\section{Experiments and Results}
\subsection{Network Structures}
Since the deliberations in the previous chapters were of a high-level nature, we will introduce the tested embodiments of the general network structures in the following subsection. Since we limited the scope of this work to residual compression techniques, we will only present the residual coding networks in detail. For motion estimation and motion transmission, we used the same components as have been used in DVC \cite{LuOX2018_DVCEndend}. In our network design of the residual coder, we closely followed the design choices from \cite{LadunePH2020_ModeNetModeSelection}. 
\newcommand{\imgEncoder}{\includegraphics[width=0.13\linewidth]{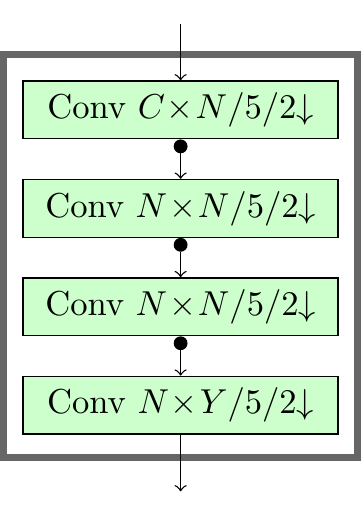}}
\newcommand{\imgDecoder}{\includegraphics[width=0.13\linewidth]{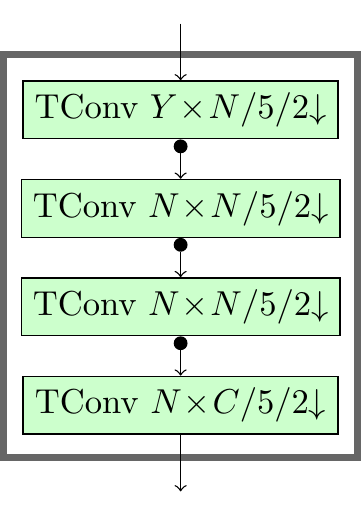}}
\newcommand{\imgEncoderP}{\includegraphics[width=0.13\linewidth]{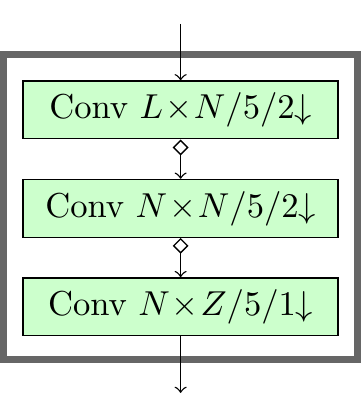}}
\newcommand{\imgDecoderP}{\includegraphics[width=0.13\linewidth]{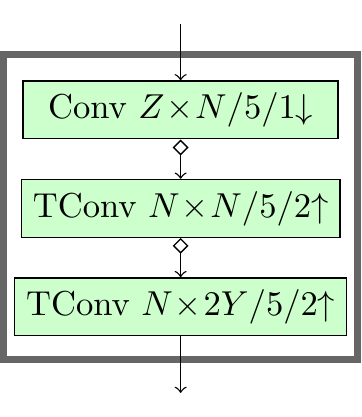}}
\newcommand{\imgGS}{\includegraphics[width=0.13\linewidth]{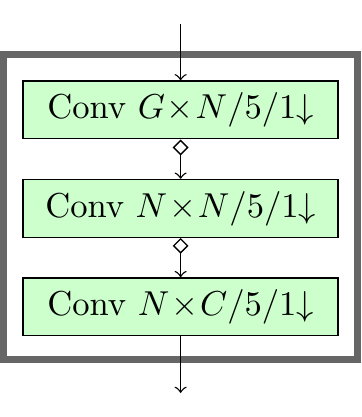}}
\newcommand{\imgGD}{\includegraphics[width=0.13\linewidth]{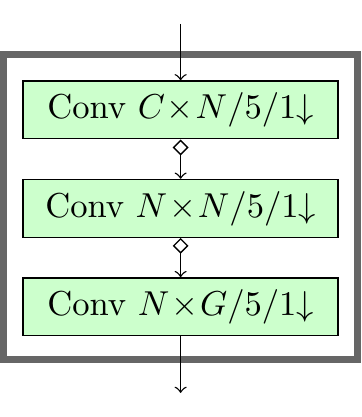}}

\newcommand{\imgDiffCoder}{\includegraphics[height=0.22\linewidth]{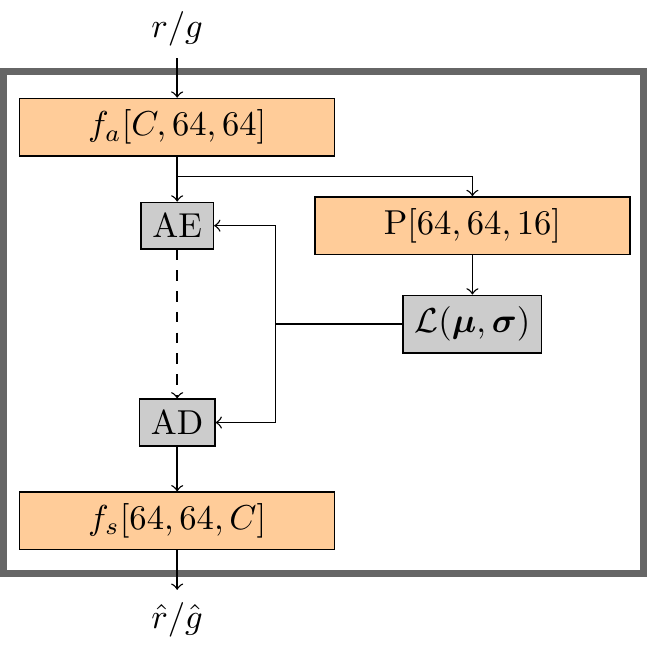}}
\newcommand{\imgCondCoder}{\includegraphics[height=0.22\linewidth]{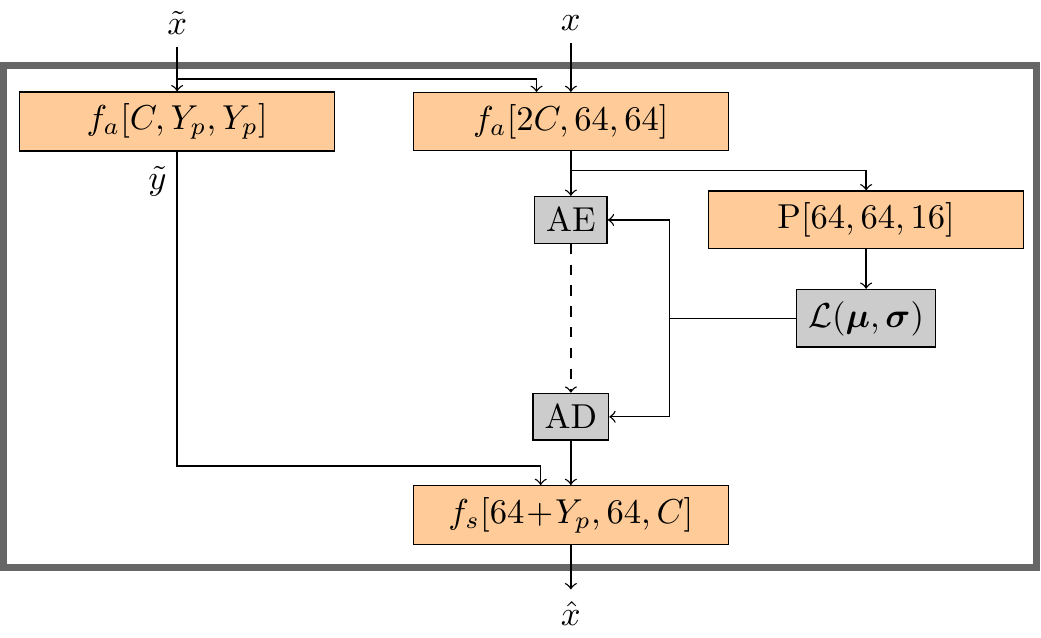}}
\newcommand{\imgHypCoder}{\includegraphics[height=0.22\linewidth]{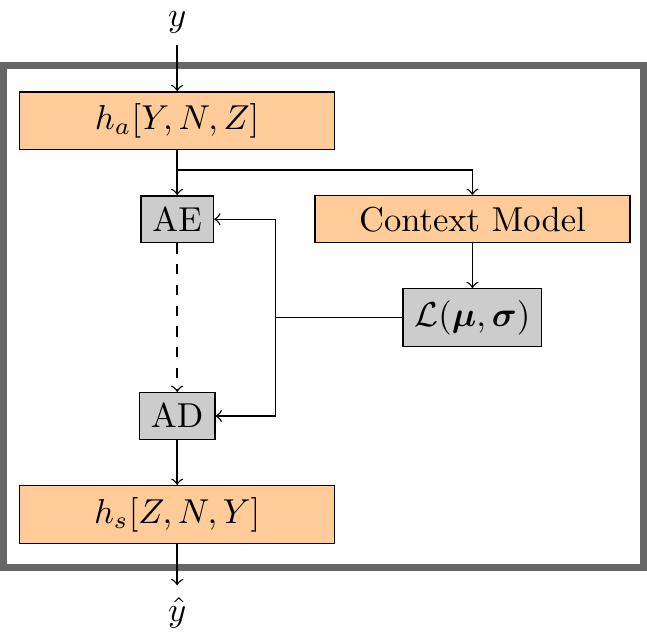}}
\newcommand{\imgxGDC}{\includegraphics[height=0.22\linewidth]{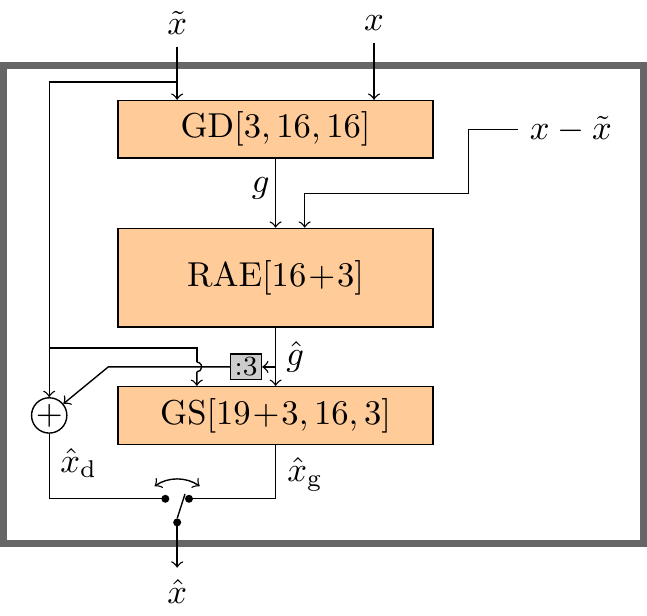}}
\newcommand{\imgGDC}{\includegraphics[height=0.22\linewidth]{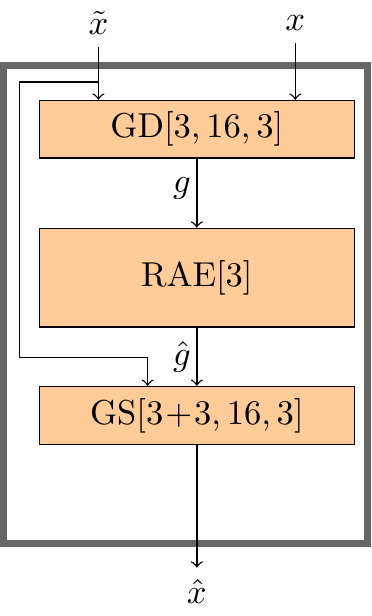}}
\newcommand{\imgBase}{\includegraphics[height=0.22\linewidth]{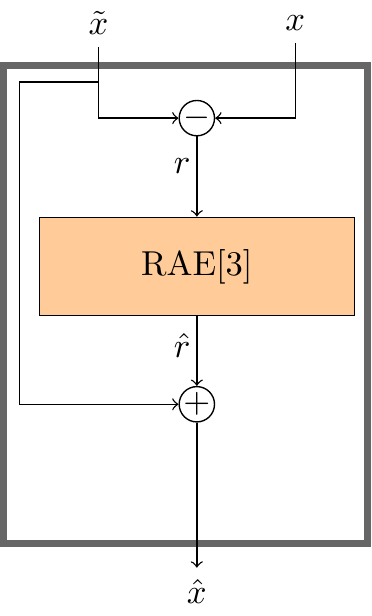}}
\newcommand{\imgCondCoderL}{\includegraphics[height=0.22\linewidth]{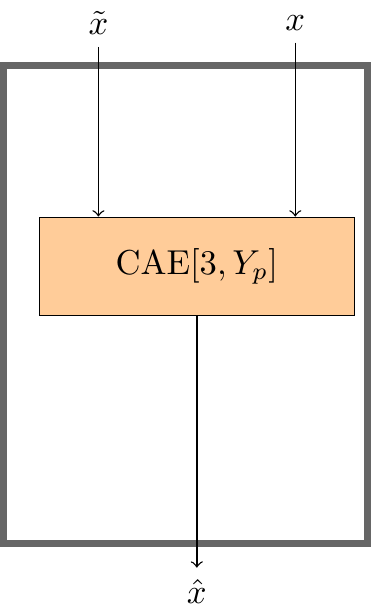}}
\begin{figure*}
	\centering
	\small
	\begin{tabular}{cccccc}
		\imgEncoder & \imgDecoder & \imgEncoderP &\imgDecoderP& \imgGD & \imgGS\\
		(a) $\encoderNet{C}{N}{Y}$ & (b) $\decoderNet{Y}{N}{C}$ & (c) $\encoderPNet{Y}{N}{Z}$ & (d) $\decoderPNet{Z}{N}{Y}$ & (e) $\GDNet{C}{N}{G}$ & (f) $\GSNet{G}{N}{C}$\\
	\end{tabular}
	\begin{tabular}{ccc}
		\imgDiffCoder & \imgCondCoder & \imgHypCoder\\
		(g) Residual Autoencoder $\resCoderNet{C}$ & (h) CodecNet $\condCoderNet{C}{Y_p}$ \cite{LadunePH2020_ModeNetModeSelection} & (i) Prior Coder $\hypCoderNet{Y}{N}{Z}$ \cite{BalleMS2018_Variationalimagecompression,MinnenBT2018_JointAutoregressiveHierarchical}
	\end{tabular}
	\begin{tabular}{cccc}
		\imgBase & \imgCondCoderL & \imgGDC & \imgxGDC \\
		(j) Residual Coder (Ref)& (k) Conditional Coder~\cite{LadunePH2020_ModeNetModeSelection} & (l) GDC & (m) xGDC\\
		\textit{DiffCoder} & \textit{CodecNet-64} & \textit{GDC} & \textit{xGDC}\\
		&\textit{CodecNet-192}&&\\
	\end{tabular}
\caption{(a)-(f) Basic building blocks of our networks. Conv $C\!\times\!N/k/s\downarrow$ denote a convolutional layer with $C$ input channels, $N$ output channels, a kernel size of $k\!\times\!k$ and a stride of $s$. TConv denotes a transposed convolution with analogous parameters. Circles denote GDN or IGDN~\cite{BalleLS2015_DensityModelingImages} non-linearities and rhombi denote PReLU activation functions. (g) Diagram of residual coder. The parameter $C$ denotes the number of input channels of the coder. (h) Diagram of a conditional autoencoder as from \cite{LadunePH2020_ModeNetModeSelection}. The additional parameter $Y_p$ denotes the number of channels in the prediction bottleneck. (i) Prior Coder similar as in \cite{BalleMS2018_Variationalimagecompression} and \cite{MinnenBT2018_JointAutoregressiveHierarchical}. The context model is used here to compress the latent representation of the prior and consists of 2 mask convolutions with 16 channels each. Two inputs into one block implicitly contain a concatenation block. The arithmetic encoder (AE) contains a quantization operation. (j)-(m) Tested coding systems. Each system receives current frame and prediction signal and outputs the reconstructed frame. The switch in (m) denotes an arbitrary combination between the two signal. The box labeled ``:3'' selects the first 3 channels out of the signal. The name in italics below denote the names we use in tables and figures. } \label{Fig:Networks}
\vspace{-0.3cm}
\end{figure*}
In Fig. \ref{Fig:Networks}, we give the network structures in detail. We compare four general types of coding systems. First, we use a standard residual coder (\textit{DiffCoder}) as baseline. We compute the difference between the input and the prediction signal and compress the resulting residual using a standard four-layer autoencoder with hyperprior and context model. We then compare two conditional coders which follow the structure of CodecNet from ~\cite{LadunePH2020_ModeNetModeSelection}. The coders only differ in the size of the prediction bottleneck. We denote them as \textit{CodecNet-64} and \textit{CodecNet-192}, where the number denotes the parameter $Y_p$ shown in the diagrams. As previously discussed, this approach includes a second encoder which generates a latent representation of the prediction signal, which is then used to reconstruct the frame. 

Next, we test our proposed GDC approach. Here, we have replaced the difference and sum with shallow three layer convolutional networks. This network is designed in a way that the residual autoencoder at the core has the same dimensionality as in the difference coder. We can therefore stabilize the training by initially training a difference coder and then tune the network with the generalized difference and sum.

Finally, we have the extended generalized difference coder xGDC. Here, we additionally encode and decode the linear residual. As discussed, this improves the feasibility of hybrid coding approaches and stabilizes the training. Since we do not need a pre-initialization of the autoencoder anymore, we are able to increase the capacity of the generalized difference to an output of 16 channels. The core autoencoder therefore has $16+3=19$ channels. 16 channels come from the generalized difference and 3 channels from the linear difference. 
\subsection{Training}
In order to exclude any influences of other components on the measured performance, we fixed the motion estimation and motion transmission modules during training. We used pre-trained models, which have been made available on GitHub\footnote{\url{https://github.com/ZhihaoHu/PyTorchVideoCompression/tree/master/DVC}}. We train our models on the CLIC 2020 training set~\cite{Mentzer_clic2020devkit}. This set consists of videos of user-generated content, which were partially coded or preprocessed in varying degree. The sequences contain text, screen content, and animated parts, making this a very diverse dataset. 

We pick image pairs out of this set and randomly crop the frames to patches of size $256\!\times\!256$. We then estimate the motion, compress the motion field and compress the residual information using one of the compression networks outlined above. We train the network using the Adam optimizer~\cite{KingmaB2015_AdamMethodStochastic} with standard parameters and a learning rate of $10^{-4}$. We train the networks on a joint rate-distortion loss function
\begin{equation}
L = D_\mathrm{MSE}(\orig, \reco) + \lambda R \label{Eq:RDO}
\end{equation}
where the multiplier $\lambda$, which determines the exact position on the rate-distortion curve, is chosen from $\lambda\in\left\{256,512,1024,2048\right\}$.

To train the xGDC, we need to find a rule which of the two signals $\xD$ or $\xG$ to use. One possibility would be to pick the one with better quality. However, this may lead to a mode collapse, particularly because $\xD$ will have a better initial performance at the beginning of the training. This would result in not training the $\xG$ path at all. We instead follow the result from (\ref{Eq:Main}), which tells us that a conditional approach works better when $I(\pred;\res)$ is large. Since a very high prediction quality leads to a small $I(\pred;\res)$, we use the quality of the prediction signal as support. When the prediction PSNR between the prediction frame and the original exceeds a threshold $\thresh=30\,\mathrm{dB}$, we choose to train $\xD$, otherwise, we choose $\xG$. We chose this threshold after preliminary experiments with different generalized difference coders.

\subsection{Test Setup}
In our experiments, we always compress one image pair at a time, where one of them is given as reference frame. We measure the total rate required to compress the second frame, including the rate to transmit the motion vectors and the residual. We perform our tests on the CLIC 2020 validation set~\cite{Mentzer_clic2020devkit}. For the tests on the CLIC set, we randomly choose 100 image pairs from each of the thirteen classes of the set. Note that this set contains many image pairs which are very similar even without motion compensation. So the residual is often very small. In all tests, we compress one frame given a prediction obtained from a previous frame. To simulate real coding conditions, we compressed the reference frame with the JPEG2000 image coder~\cite{ITUTI2004_JPEG2000Image} to an average image quality of 35\,dB PSNR.
\subsection{Results}
\begin{figure}
	\includegraphics[width=\linewidth]{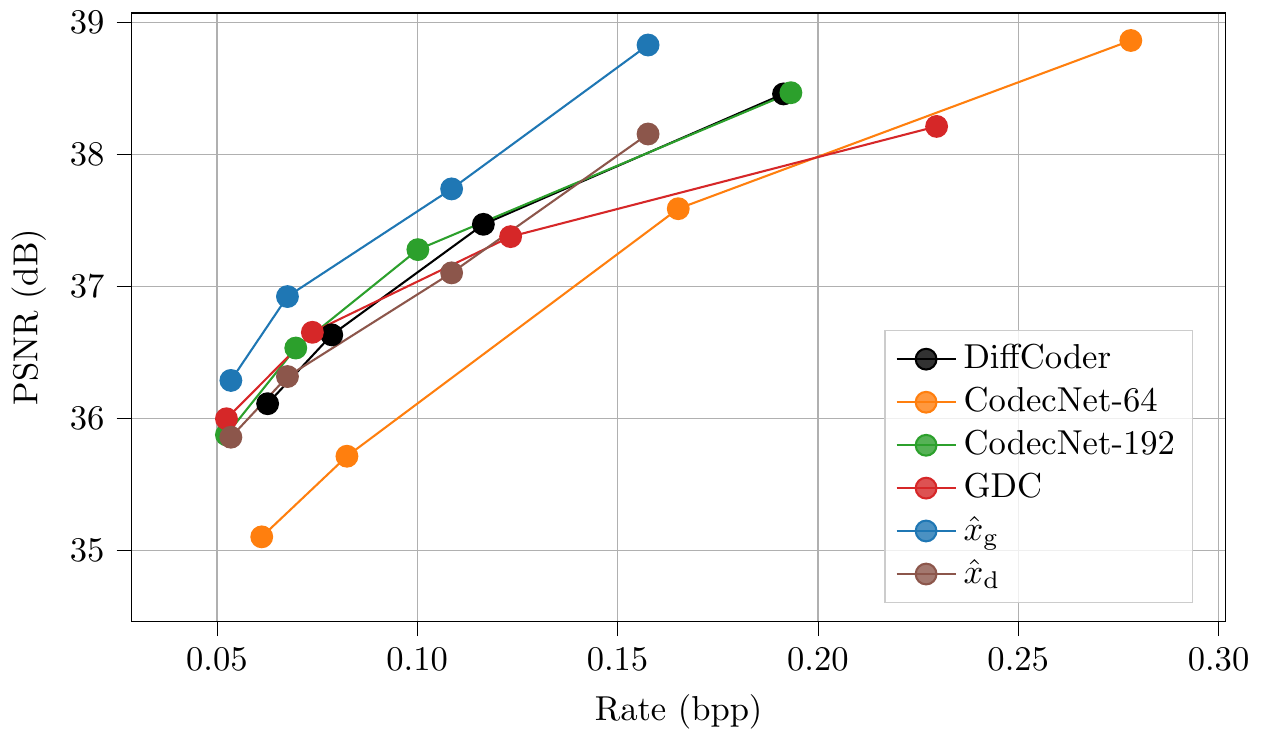}
	\caption{Rate-distortion curve for the CLIC validation set for the different tested methods. The reference image was compressed with JPEG2000.}\label{Fig:RDPlotCLIC}\vspace{-0.3cm}
\end{figure}

For a first experiment, we compare our coder on the CLIC dataset. We show the results in Fig.~\ref{Fig:RDPlotCLIC}. Here, we first see that the standard conditional coders CodecNet-64 performs worse than the difference coder. Note that originally CodecNet was proposed together with a skip mode network. A detailed visual analysis of the result of this network showed that in areas of very good prediction, the network has yield losses. Those are particularly the regions targeted by a skip mode. Examining a skip mode is not within the scope of this paper. We rather aim to examine and improve the performance and robustness of different conditional coding approaches in a general scenario. 

We tested two variants of CodecNet which only differ in the size of the prediction bottleneck, i.e. the size of the prediction latent space. The orange curve which depicts the behavior with a latent space with 64 channels is always below the green curve which shows the case with a latent space with 192 channels. This is further evidence for the theoretical deliberations from Section~\ref{Sec:Theory}. 

When we look at the generalized difference coder GDC, we see that its performance is similar to the performance of the difference coder. In the plot we also show the individual components of xGDC. We see that $\xG$ performs better than all other tested methods and that $\xD$ matches the performance of the difference coder. We designed xGDC with hybridization in mind. We therefore also want to analyze the performance in hybrid approaches. To that end, we compare hybrid versions of xGDC, GDC and CodecNet-192. For xGDC we switch between $\xG$ and $\xD$, for GDC and CodecNet-192 we switch between the conditional coding method and the difference coder. For each frame, we test both methods and decide on the final selected coder according to the rate-distortion loss function (\ref{Eq:RDO}). Note that this kind of frame-wise or tile-wise hybridization is possible for all coding methods but is not scalable. For larger frames or tiles, the different characteristics can no longer be exploited in different image regions. Only xGDC is able to switch methods on a finer granularity since both candidates are decoded from the same latent space. For better comparability, we test both tile-wise switching and spatially variable switching methods. 

\begin{figure}
	\includegraphics[width=\linewidth]{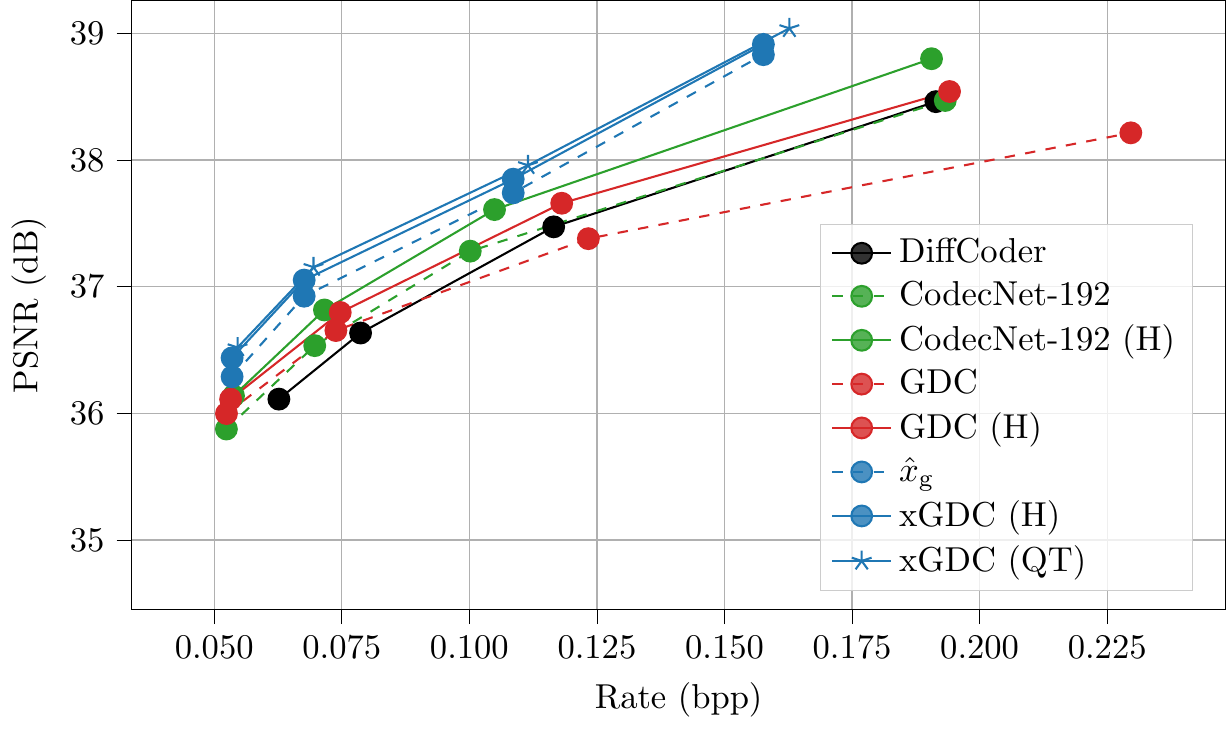}
	\caption{Rate-distortion curve for the CLIC validation set for the different tested methods. We compare CodecNet-192, GDC and xGDC (dashed lines) with their respective hybrid methods (solid lines).}\label{Fig:RDPlotCLICH}\vspace{-0.6cm}
\end{figure}

In Fig.~\ref{Fig:RDPlotCLICH} we observe that hybridization of conditional coding and residual coding can greatly increase the coding efficiency. This demonstrates that there are distinct images for which conditional coders perform better. When we look at the hybrid methods, we see that both CodecNet-192 and GDC perform better than the reference coder. Using the Bj\o ntegaard delta rate~\cite{Bjontegaard2001_CalculationaveragePSNR} (BD-Rate), we can compute that we save 15.9\% rate compared to the difference coder with CodecNet-192 and 9.3\% with GDC. Note that CodecNet-192 comes with a considerable increase of parameters. This effect is particularly strong on the decoder side, where a large network with four layers with 192 channels each has to be executed. This network alone has about 2.8 million additional parameters. In contrast, the generalized difference coder adds 2 layers with 16 channels and 1 layer with 3 channels to encoder and decoder each. This amounts to a total of only 20140 additional parameters. This demonstrates that the gain can be achieved much easier with the generalized difference coder. 

\begin{figure*}[t!]
	\begin{tabular}{c@{\hspace{0.02\linewidth}}c}
		\includegraphics[width=0.49\linewidth]{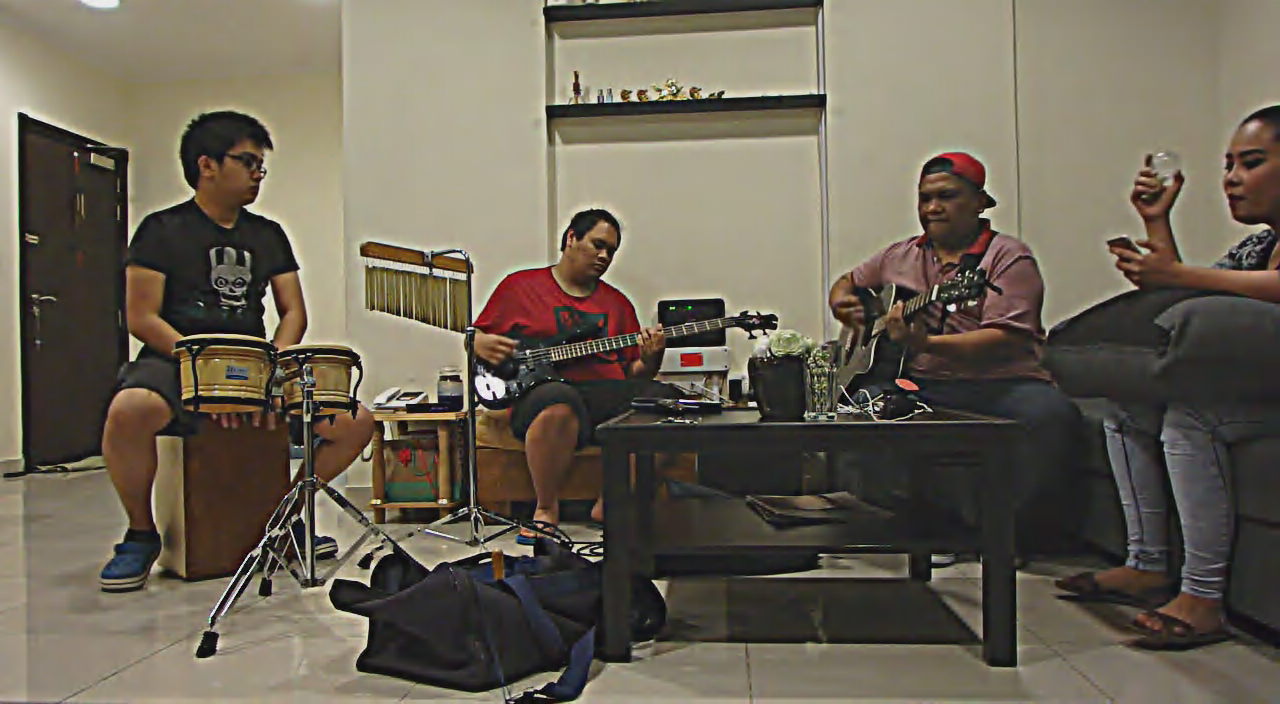} & \includegraphics[width=0.49\linewidth]{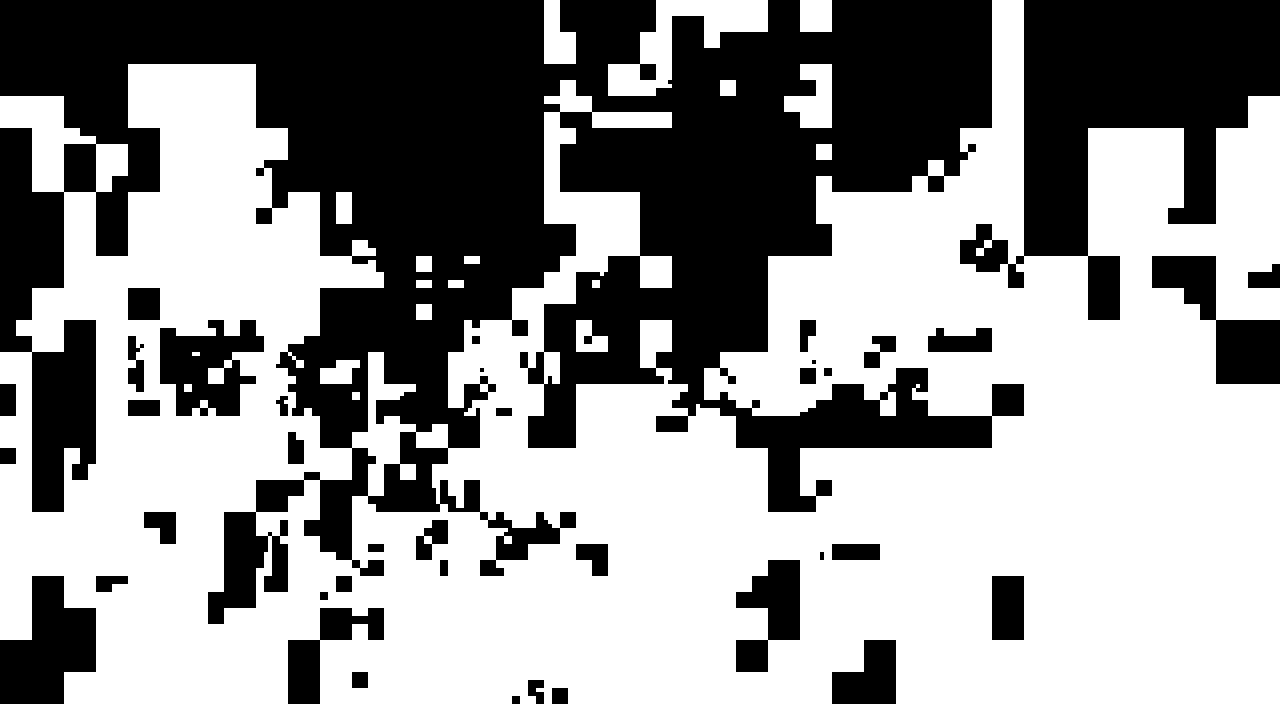}\\
		xGDC-QT & Quad Tree Mask
	\end{tabular}
	\begin{tabular}{c@{\hspace{0.02\linewidth}}c@{\hspace{0.02\linewidth}}c@{\hspace{0.02\linewidth}}c@{\hspace{0.02\linewidth}}c@{\hspace{0.02\linewidth}}c@{\hspace{0.02\linewidth}}c}
	\includegraphics[width=0.125\linewidth, trim=580 250 270 200, clip]{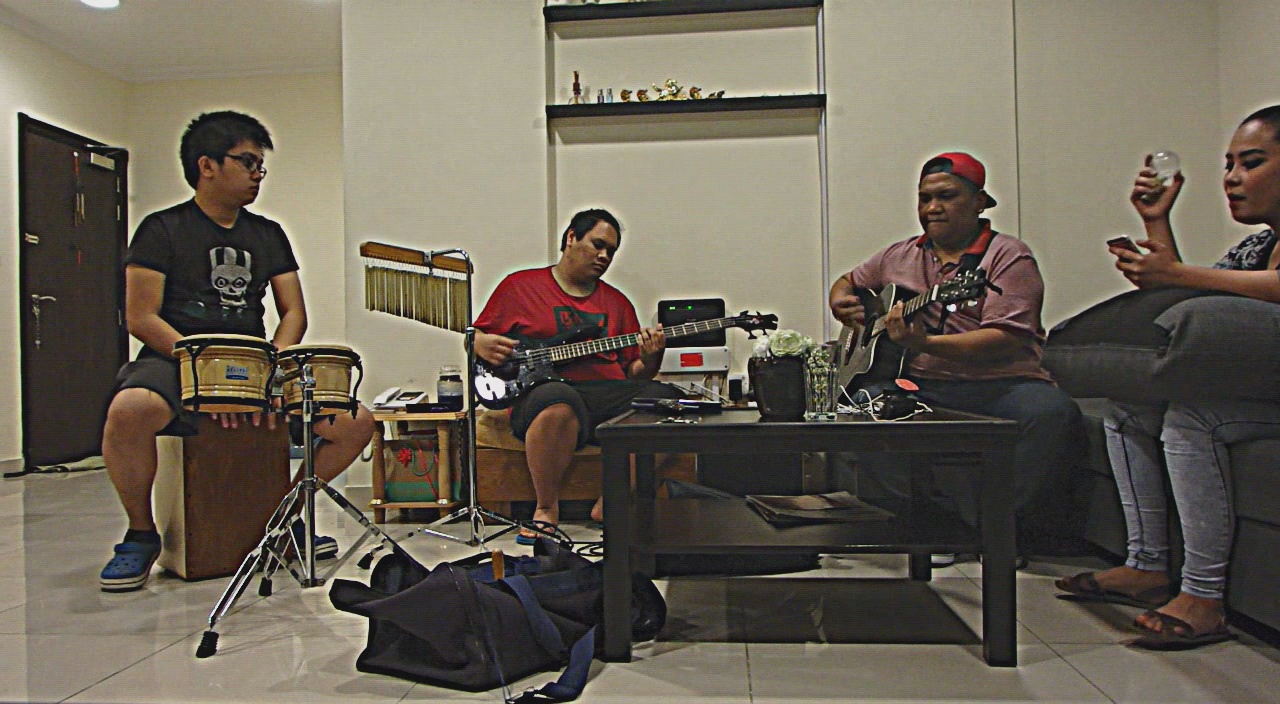} & 
		\includegraphics[width=0.125\linewidth, trim=580 250 270 200, clip]{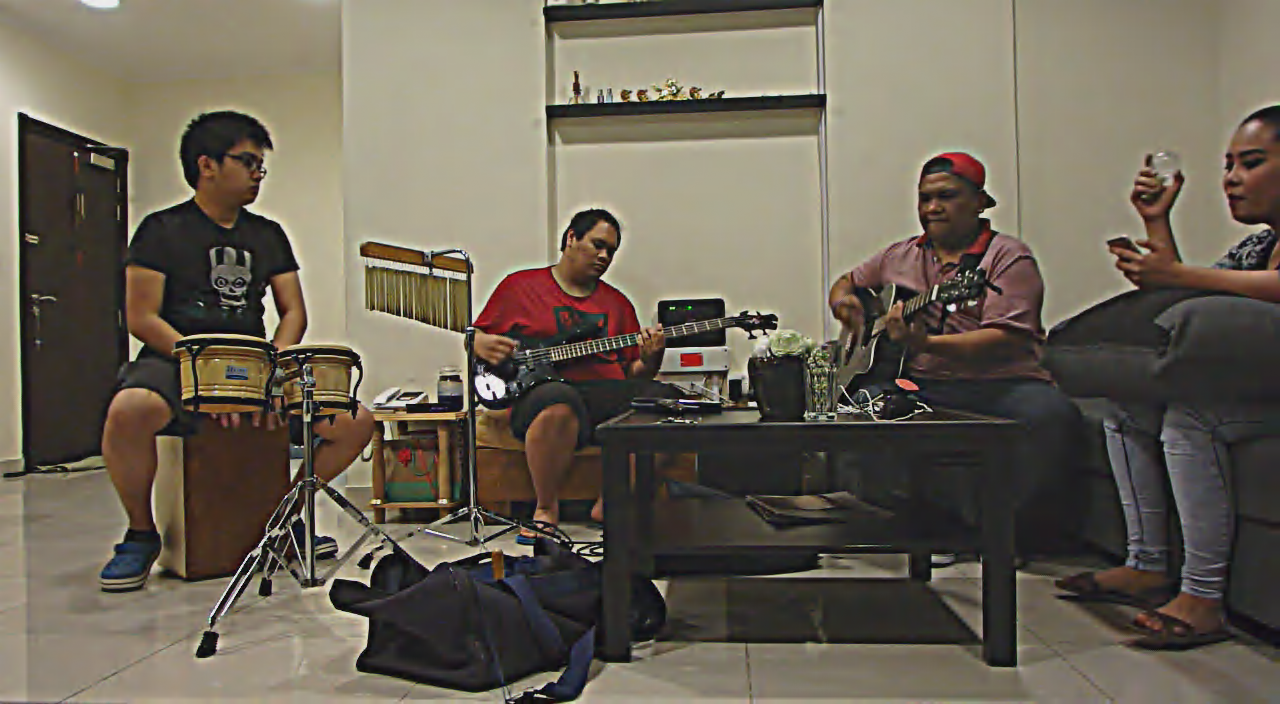}  & 
		\includegraphics[width=0.125\linewidth, trim=580 250 270 200, clip]{Figures/CondCoderGenDiffInitV3x2_C30_QT_CoverSongF_21_1024_0_02_reco.png} & 
		\includegraphics[width=0.125\linewidth, trim=580 250 270 200, clip]{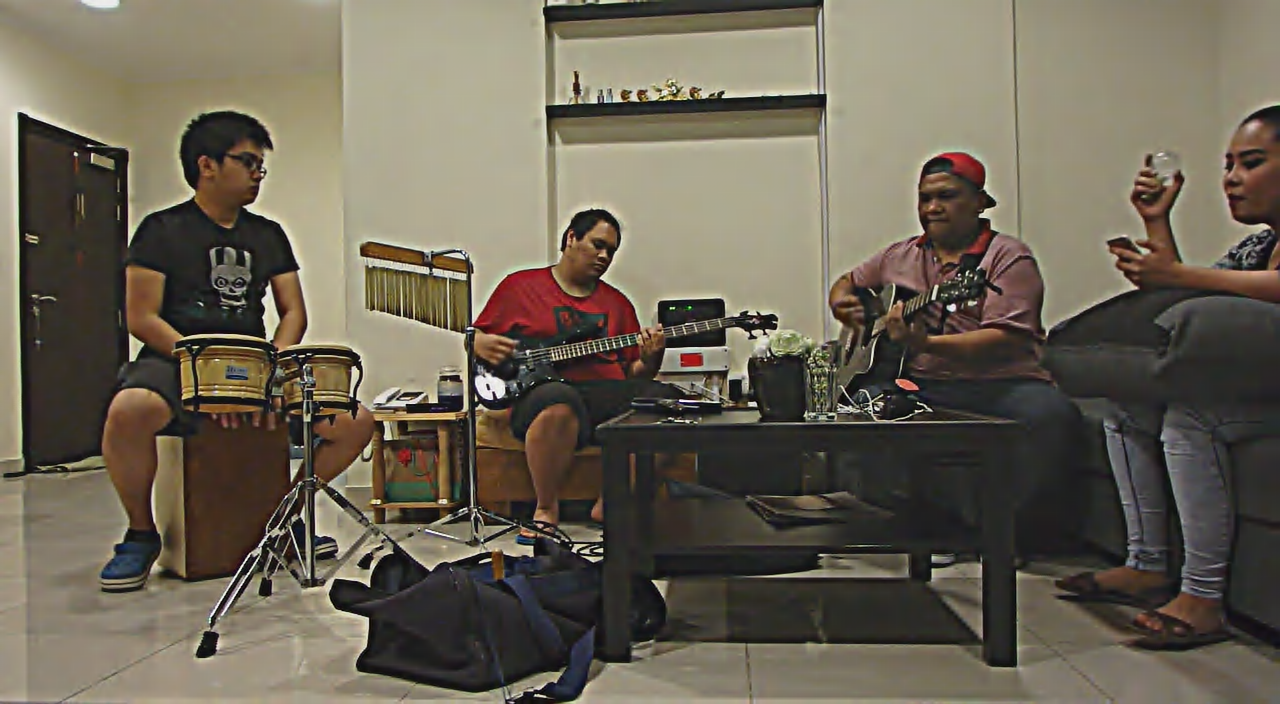} & 
		\includegraphics[width=0.125\linewidth, trim=580 250 270 200, clip]{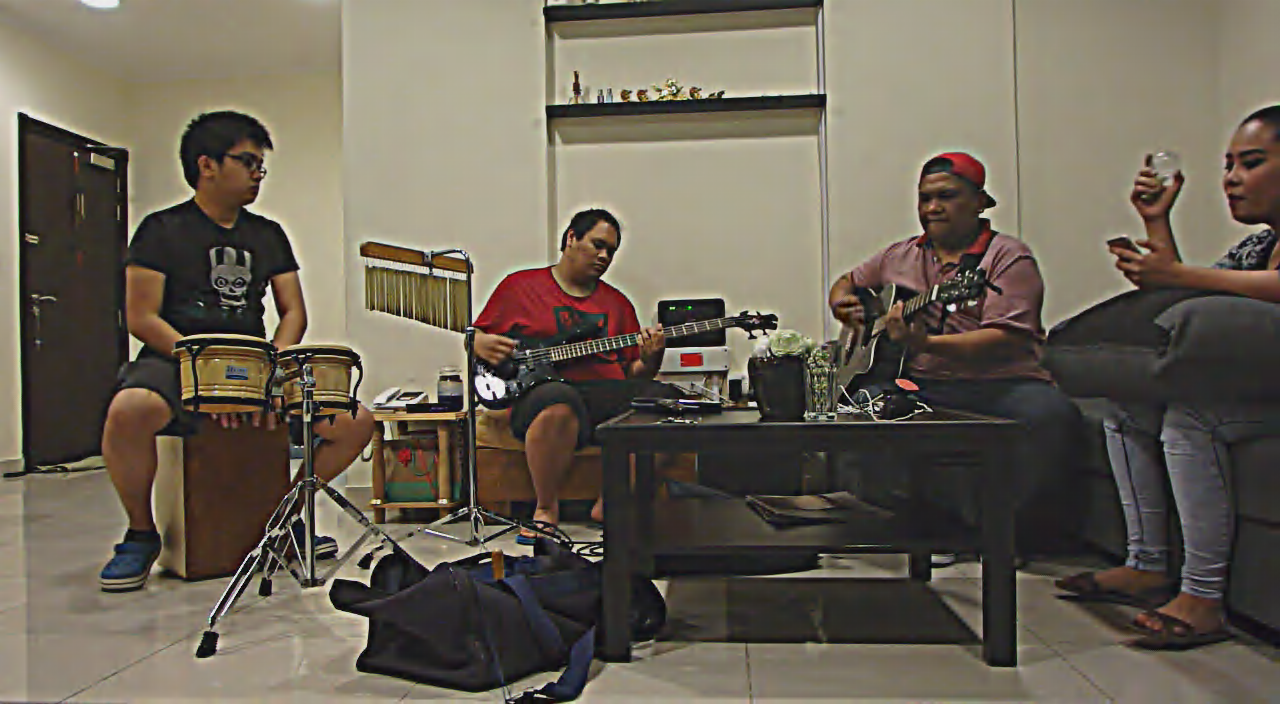}  
		& \includegraphics[width=0.125\linewidth, trim=580 250 270 200, clip]{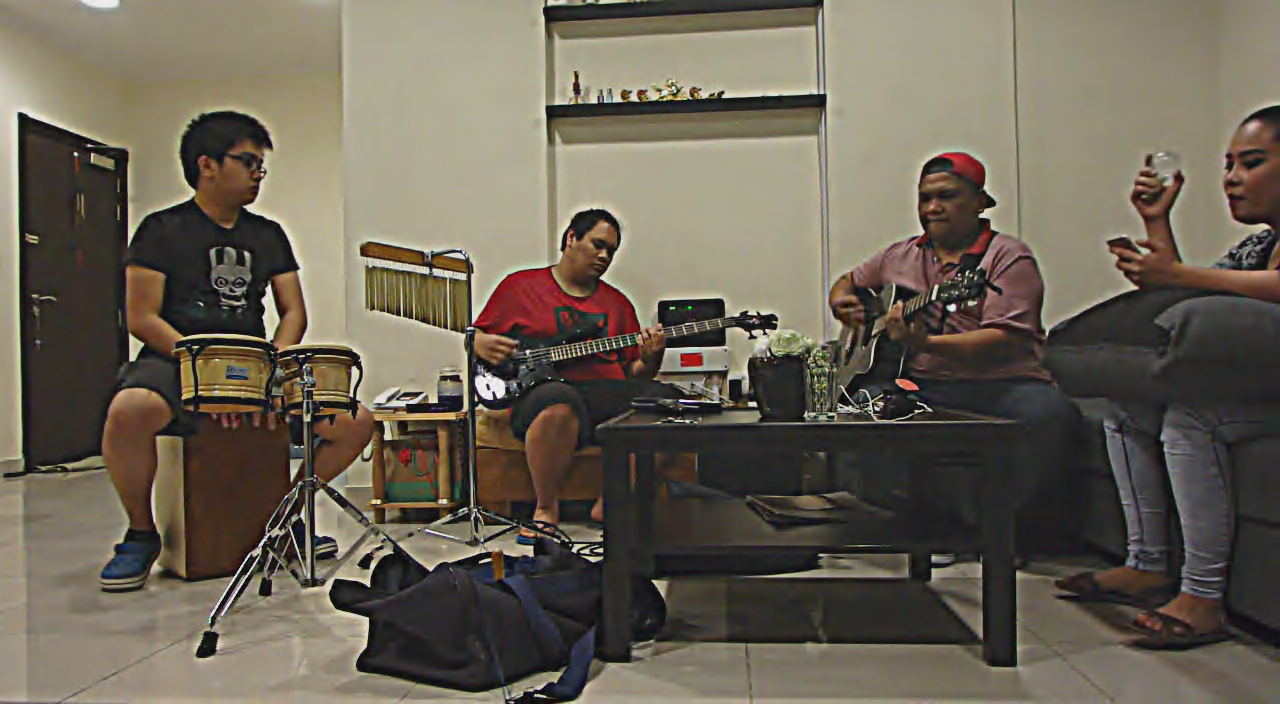}
		& \includegraphics[width=0.125\linewidth, trim=580 250 270 200, clip]{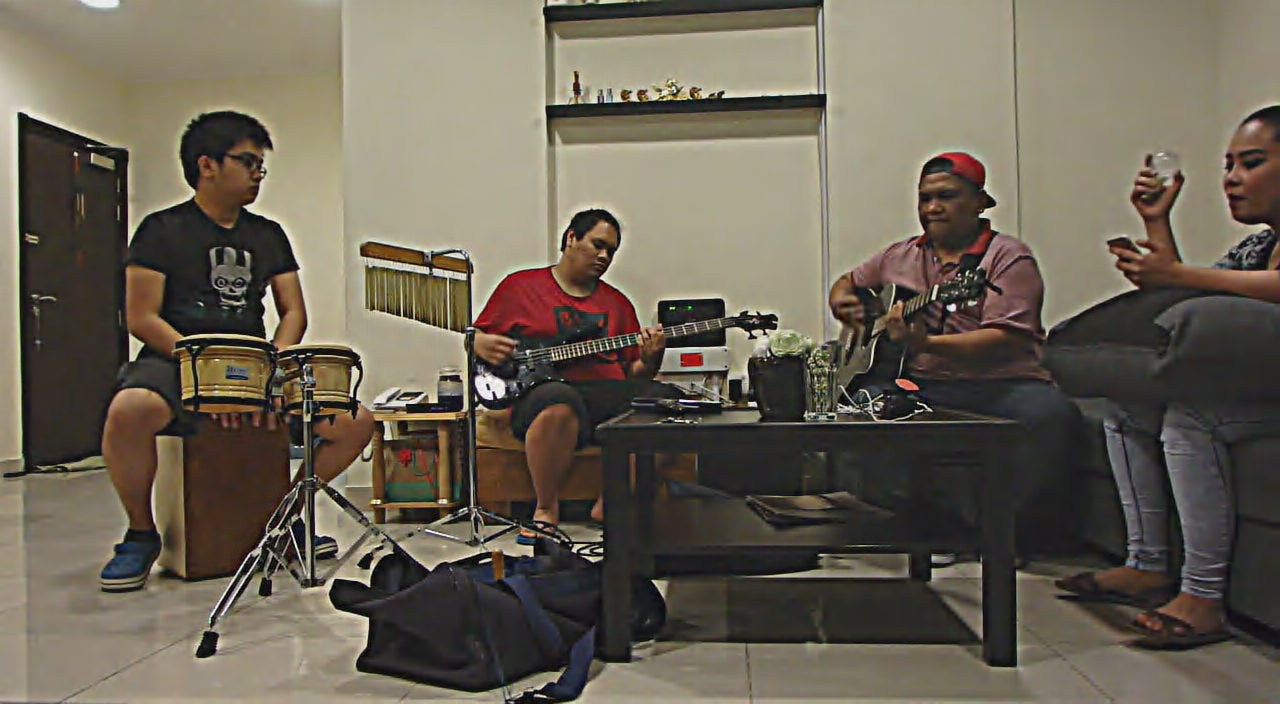}\\
		Original&Prediction&xGDC-QT&$\xG$&$\xD$&DiffCoder&CodecNet-192\\		
	\end{tabular}
	\caption{Visual example of the tested compression algorithms. In the top row, we show the full reconstructed frame after using a quadtree hybridization with xGDC. On the right, we show the corresponding mask. In white areas $\xG$ was chosen, in black areas, $\xD$ was chosen. Below, we compare several reconstructed signals on one zoomed-in excerpt. The images were coded at around 0.04\,bpp. The images are best viewed digitally on a well lit screen.}\label{Fig:VisualExample}
	\vspace{-0.3cm}
\end{figure*}

Fig. \ref{Fig:RDPlotCLICH} shows that also in the hybrid scenario xGDC outperforms all other methods, saving 26.1\% rate on average compared to the difference coder. Note that $\xG$ alone performs better than the hybrid versions of CodecNet-192 and GDC, saving 21.9\%. This clearly shows the larger robustness of xGDC against different image characteristics. Naturally, the hybridization gain is smaller for xGDC, since the $\xG$ already clearly outperforms the difference coder. xGDC furthermore enables us to perform hybridization on a finer level. To find rate-distortion-optimal partitionings, we employ a quad-tree-search. Similar techniques are used in HEVC for block partitioning~\cite{SullivanOH2012_OverviewHighEfficiency}. We test blocks ranging from $256\!\times\!256$ to $4\!\times\!4$. This method only costs very little side information, since the quad-tree structure can be transmitted efficiently. To transmit the selected signal, we additionally need 1 bit per block. All this side-information is taken into account during the quad-tree-search. When performing this search, we can achieve additional gains, yielding overall rate savings of 27.8\% against the reference coder. Also note the small complexity overhead of xGDC. In this method, we only increase the number of parameters by 6.6\% compared to the reference model. In contrast, CodecNet-64 needs 25\% more parameters and CodecNet-192 increases the number of parameters by 358\%. We summarize the results in Tab. \ref{Tab:Results}.

\begin{table}
	\small
	\begin{tabular}{lccc}
		\toprule
		&CodecNet-192 (H) & GDC (H) & xGDC (QT)\\
		\midrule
		BDR & -15.9\% & -9.3\% & -27.8\%\\
		Complexity & +358\% &+1.6\% & +6.7\%\\
		\bottomrule
	\end{tabular}
\caption{Summary of rate savings and complexity overhead of the difference approaches. Rate savings are given in Bj\o ntegaard delta rate compared to the difference coder. Negative values denote rate savings. The complexity is given as additional number of parameters relative to the difference coder.}\vspace{-0.5cm}\label{Tab:Results}
\end{table}

Finally, we want to demonstrate the performance of our method on a visual example. In Fig. \ref{Fig:VisualExample}, we show one image after compression with xGDC using a quad-tree partitioning. We also show the mask. The shape of the mask confirms our assumptions. We see that $\xG$ is chosen where the prediction signal is worse. This is the case in highly structured areas and in areas with large motion, e.g., the hand of the guitar player. We choose this area to compare the methods in more detail. We indeed see a large amount of motion blur in the prediction signal and small artifacts on the top of the fingers. We see that both $\xD$ and the difference coder are not able to completely compensate the artifacts. We also see that the hand is more blurry in $\xD$. Also compared to CodecNet-192, xGDC can preserve more details.

\section{Conclusion}
In this paper, we presented our research on conditional coding for inter frame coding in video compression. Following theoretical considerations, we propose the generalized difference coder (GDC). By designing our coder in a way that it avoids bottlenecks in the prediction path, we are able to considerably reduce the number of additional parameters compared to comparable conditional autoencoders from the literature. Using small networks, we are able to include the prediction signal in the decoding process without a prior transformation in a latent representation.

Furthermore, we extend the approach to enable efficient hybridization between conditional and linear approaches. The extended generalized difference coder (xGDC) combines the strengths of conditional and residual coding. We jointly compress the conditional information and the residual. That way, we can decode two candidate frames from the same latent space and switch between them as needed on arbitrarily fine granularities. This approach only requires 6.6\% more parameters compared to a difference coder and has the additional advantage that the residual adds additional context in the coding process which can be exploited during the conditional coding. Therefore, even without any hybridization, the extended generalized difference coder outperforms both residual coders and other conditional coders with very small complexity overhead. This result demonstrates that concepts from traditional coders should not simply be replaced by neural-network-based counterparts, but rather be extended. 

In this paper, we limited ourselves to the problem of residual compression to provide a detailed analysis of conditional coding concepts in this context. In future work, we plan to include the concepts in full video coders and further examine the interaction between the components, such as different intra coding methods or inter skip modes. Since the presented concepts are of very generic nature and achieve large gains over other methods, we expect large portions of the gain to transfer to different scenarios.
\clearpage
{\small
\bibliographystyle{ieee_fullname}

}

\end{document}